\documentclass[preprint,tightenlines,aps,floats,nofootinbib]{revtex4}  
  
\usepackage{epsf}

\newcommand{\notE}{\ \hbox{{$E$}\kern-.60em\hbox{/}}}
\newcommand{\notp}{\ \hbox{{$p$}\kern-.43em\hbox{/}}}
\def\D0{\mbox{D\O}}

\includegraphics{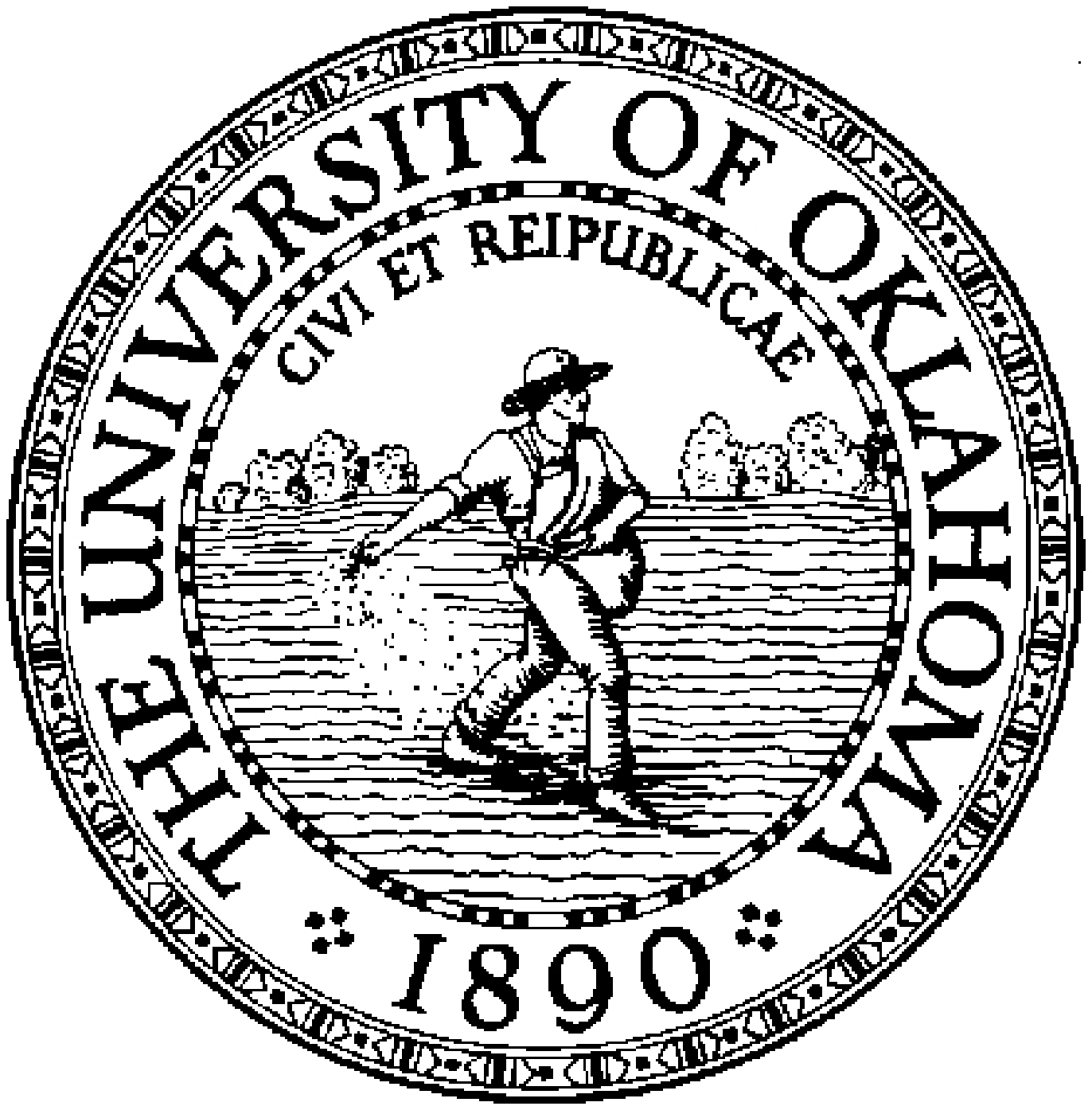}

\preprint{\font\fortssbx=cmssbx10 scaled \magstep2
\hbox to \hsize{
\hskip1.8in 
\hbox{\fortssbx The University of Oklahoma}
\hskip0.8in $\vcenter{
                      \hbox{\bf OKHEP-04-03}
                      \hbox{\bf hep-ph/0411331}
                      \hbox{November 2004}}$ }
}
  
\begin{document}  

  
\title{\vspace*{.75in}
Detecting a Higgs Pseudoscalar with a Z Boson Produced in 
Bottom Quark Fusion}

\author{
Chung Kao\footnote{E-mail address: Kao@physics.ou.edu} and 
Shankar Sachithanandam\footnote{E-mail address: Shankar@physics.ou.edu}}
  
\affiliation{  
Department of Physics and Astronomy, University of Oklahoma,  
Norman,  OK 73019, USA
\vspace*{.5in}}
  
     
\thispagestyle{empty}  


\begin{abstract}  

We investigate the prospects of detecting a Higgs pseudoscalar ($A^0$) 
in association with a $Z$ gauge boson produced from bottom quark fusion 
($b\bar{b} \to ZA^0$) at the CERN Large Hadron Collider (LHC).
A general two Higgs doublet model and the minimal supersymmetric standard 
model are adopted to study the discovery potential of 
$pp \to ZA^0 \to \ell \bar{\ell} b\bar{b} +X, \ell = e$ or $\mu$,   
via $b\bar{b} \to ZA^0$ with physics backgrounds and realistic cuts. 
Promising results are found for $m_A \alt 400$ GeV in a general two Higgs 
doublet model when the heavier Higgs scalar ($H^0$) can decay into 
a $Z$ boson and a Higgs pseudoscalar ($H^0 \to ZA^0$).
We compare the production rates from bottom quark fusion ($b\bar{b} \to ZA^0$) 
and gluon fusion ($gg \to ZA^0$), and find that they are complementary 
processes to produce $ZA^0$ in hadron collisions. 
While gluon fusion is the major source for producing a Higgs pseudoscalar 
associated with a $Z$ boson at the LHC for $\tan\beta \alt 10$, bottom quark 
fusion can make dominant contributions for $\tan\beta \agt 10$.

\end{abstract}  
  

\maketitle

\newpage  

\section{INTRODUCTION}

In the Standard Model (SM), the Higgs mechanism requires only one Higgs 
doublet to generate masses for fermions and gauge bosons. It leads to the 
appearance of a neutral CP-even Higgs scalar after spontaneous electroweak 
symmetry breaking. The LEP2 experiments have established a lower bound of 
114.4~GeV \cite{LEP2a} for the SM Higgs boson mass at 95\% confidence level.

A two Higgs doublet model (2HDM) \cite{guide} has Higgs doublets 
$\phi_1$ and $\phi_2$ with the vacuum expectation values $v_1$ and $v_2$.  
There are five physical Higgs bosons: a pair of singly charged Higgs bosons 
$H^{\pm}$, two neutral CP-even scalars $H^0$ (heavier) and $h^0$ (lighter), 
and a neutral CP-odd pseudoscalar $A^0$. The couplings of the Higgs bosons 
to fermions and gauge bosons depend on the ratio of vacuum expectation values 
($\tan\beta \equiv v_2/v_1$) and a mixing angle ($\alpha_H$) between the 
weak and mass eigenstates of the neutral scalars.

The supersymmetry between a boson and a fermion preserves the elementary 
nature of Higgs bosons. The minimal supersymmetric standard model (MSSM) 
\cite{MSSM} requires two Higgs doublets $\phi_1$ and $\phi_2$ coupling to 
fermions with weak isospin $t_3 = -1/2$ and $t_3 = +1/2$ respectively, 
to generate masses for fermions and gauge bosons and to cancel triangle 
anomalies associated with the fermionic partners of the Higgs bosons. 
The Higgs potential is constrained by supersymmetry such that all tree-level 
Higgs boson masses and couplings are determined by just two independent 
parameters, commonly chosen to be the mass of the CP-odd pseudoscalar ($m_A$) 
and $\tan\beta$. The mixing angle $\alpha_H$ between the neutral scalars is 
often chosen to be negative ($-\pi/2 \leq \alpha_H \leq  0$). 
The LEP2 collaborations have set a lower bound of 91~GeV and 91.9 GeV 
\cite{LEP2b} for the $m_h$ and the $m_A$, respectively. 

Extensive studies have been made for the detection of a heavier 
MSSM Higgs boson ($\phi^0 = H^0$ or $A^0$) at the CERN LHC 
\cite{HGG,Neutral,HZZ,Kunszt,HXX2,AZh,Nikita,Sally1,Sally2,CMS,ATLAS1,ATLAS2}. 
For $\tan\beta \alt 5$, $A^0 \to \gamma\gamma$, $H^0 \to ZZ^* \to 4\ell$,  
and $A^0,H^0 \to t\bar{t}$ are possible discovery channels.
In the SM, $H \to ZZ^* \to 4\ell$ offers great promise for 
160 GeV $\alt m_H \alt 1$ TeV \cite{CMS,ATLAS1,ATLAS2} with physics background 
from $q\bar{q} \to ZZ$ \cite{qqzz} and $gg \to ZZ$ \cite{ggzz,Glover}. 
However, the heavy scalar in the MSSM is observable via $H^0 \to ZZ^*$ only 
for $\tan\beta \alt 10$ and $m_A \alt 350$ GeV \cite{HZZ}.  
The detection modes $A^0 \to Zh^0 \to l^+l^- \tau\bar{\tau}$ \cite{AZh} 
or $l^+l^- b\bar{b}$ \cite{AZh,CMS,ATLAS2} 
and $H^0 \to h^0 h^0 \to b\bar{b} \gamma\gamma$ \cite{ATLAS2} 
may be promising channels for simultaneous discovery of 
two Higgs bosons in the MSSM. 
For large values of $\tan\beta$, 
$\phi^0 \to \mu\bar{\mu}$ \cite{Nikita,Sally1,Sally2,CMS,ATLAS2}, and 
$\phi^0 \to \tau\bar{\tau}$ \cite{Kunszt,CMS,ATLAS1,ATLAS2},  
are promising discovery channels for the $A^0$ and the $H^0$. 
In some regions of parameter space, the rates for Higgs boson
decays to neutralinos ($H^0,A^0 \to \chi^0_2 \chi^0_2$) are dominant and 
they might open up new promising modes for Higgs detection \cite{HXX2}.

In two Higgs doublet models, there are two complementary channels 
to search for a Higgs scalar and a Higgs pseudoscalar simultaneously: 
(i) $A^0 \to Zh^0$ \cite{AZh,CMS,ATLAS2} 
with a coupling proportional to $\cos(\beta-\alpha_H)$ and 
(ii) $H^0 \to ZA^0$ 
with a coupling proportional to $\sin(\beta-\alpha_H)$. 
At the LHC, gluon fusion can be a significant source to produce 
a Higgs pseudoscalar ($A^0$) and a $Z$ boson ($gg \to ZA^0$)
via triangle and box diagrams with the third generation quarks
\cite{ggza1,Yin:2002sq}. The top quark loop diagrams make dominant  
contribution to $gg \to ZA^0$ for $\tan\beta \alt 10$ \cite{ggza1}.

In the MSSM and a 2HDM with Model II Yukawa interactions \cite{Model2} for 
the Higgs bosons and fermions, one Higgs doublet ($\phi_1$) couples to 
down-type quarks and charged leptons while another doublet ($\phi_2$) couples 
to up-type quarks and neutrinos. The $A^0 b\bar{b}$ coupling is proportional 
to $\tan\beta$ and this coupling can be greatly enhanced by a large value 
of $\tan\beta$. In addition, the Higgs pseudoscalar does not couple to gauge 
boson pairs at the tree level. Therefore, $A^0 \to b\bar{b}$ is the dominant 
decay channel for $\tan\beta \agt 10$ or for $m_A \alt m_Z +m_h \sim 210$ GeV 
with $\tan\beta \sim 2$. 
Recently, we demonstrated that gluon fusion could be a promising production 
mechanism to detect $pp \to ZA^0 \to \ell\bar{\ell}  b\bar{b} +X$ 
via $gg \to ZA^0$ for $\tan\beta \sim 2$ and $m_A \alt 260$ GeV \cite{ggza2}. 

In this article, the prospects of the search for a Higgs pseudoscalar ($A^0$) 
associated with a $Z$ boson produced are investigated.  
We study the discovery potential of 
$pp \to ZA^0 \to \ell\bar{\ell} b\bar{b} +X$) via bottom quark fusion 
($b\bar{b} \to ZA^0$) at the LHC.
The production cross sections of $ZA^0$ at the LHC 
in a two Higgs doublet model and the MSSM are discussed in Section II.
The dominant physics backgrounds from production of 
$\ell\bar{\ell}b\bar{b}$ and $W^+ W^- b\bar{b}$ are presented in Section III.
The discovery potential of $ZA^0 \to \ell\bar{\ell}b\bar{b}$ is discussed 
in Section IV. Conclusions are drawn in Section V.

\section{THE PRODUCTION CROSS SECTIONS}

We calculate the cross section for $pp \to ZA^0 +X$ via $b\bar{b} \to ZA^0$ 
in a two Higgs doublet model and the minimal supersymmetric standard model 
with Model II Yukawa interactions for the Higgs bosons and fermions. 
The parton distribution functions of CTEQ6L1 \cite{CTEQ6} 
are employed to evaluate the cross section for 
$pp \to ZA^0 \to \ell\bar{\ell} b\bar{b} +X$ 
with the Higgs production cross section $\sigma(pp \to ZA^0 +X)$ 
multiplied by the branching fractions of $Z \to \ell\bar{\ell}$ 
and $A^0 \to b\bar{b}$. 
In the Yukawa couplings of $\phi^0 b\bar{b}$ ($\phi^0 = A^0, H^0, h^0$), 
the bottom quark mass is chosen to be the next-to-leading order (NLO) 
running mass $m_b(m_A)$ \cite{bmass}, which is calculated with 
$m_b({\rm pole}) = 4.7$ GeV and the NLO evolution of the strong coupling 
\cite{alphas}.

In Figure 1, we present the Feynman diagrams for $b\bar{b} \to ZA^0$.
The s-channel diagrams contain  
the heavy Higgs scalar $H^0$ [$g_{AZH} \propto \sin(\beta-\alpha_H)$] and 
the light Higgs scalar $h^0$ [$g_{AZh} \propto \cos(\beta-\alpha_H)$]  
in the intermediate state. Therefore, this discovery channel might provide 
a good opportunity to measure the couplings of $ZH^0 A^0$ and $Zh^0 A^0$. 
The $t$ and $u$ channel diagrams are proportional to $\tan\beta$. 
When the $t$ and $u$ channel diagrams dominate, the cross section 
of $b\bar{b} \to ZA^0$ can be enhanced by a large value of $\tan^2\beta$. 
We have checked the relevant couplings with the unitarity 
condition for the total amplitude. At very high energy, the longitudinal 
polarization vector ($\epsilon_\mu$) of the $Z$ boson can be expressed as 
the momentum vector over $Z$ mass ($p_\mu/M_Z$). Then unitarity requires 
cancellation among the $s$, $t$, and $u$ channel diagrams at high energy.

Figure 2 presents the cross section of 
$pp \to ZA^0 \to \ell\bar{\ell} b\bar{b} +X$ via $b\bar{b} \to ZA^0$ 
as a function of $\tan\beta$ in a general two Higgs doublet model and 
the minimal supersymmetric standard model. 
We have chosen $m_H = m_A + 100$ GeV, $m_h = 120$ GeV, 
and $\alpha_H = -\pi/4$ for the general 2HDM.
It is clear that the cross section in a 2HDM can be significantly larger than 
that in the the MSSM when the $H^0$ can decay into $ZA^0$ with 
$m_H > m_A +M_Z$ since $m_H$ and $\alpha_H$ are free parameters in a 2HDM.
In the MSSM with $\tan\beta \agt 10$, $m_A$ and $m_h$ are very close to each 
other for $m_A \alt$ 125 GeV, while $m_A$ and $m_H$ are almost degenerate 
when $m_A \agt$ 125 GeV \cite{Nikita}. 
Therefore, the decay $H^0 \to ZA^0$ is kinematically inaccessible in the MSSM. 
The cross section for gluon fusion alone is also presented in this figure.
While gluon fusion is the major source for producing $ZA^0$ with 
$\tan\beta \alt 10$, bottom quark fusion can make dominant contribution 
for $\tan\beta \agt 10$. 

The Higgs production rate via bottom quark fusion is very sensitive to the 
choice of factorization scale ($\mu_F$) \cite{Maltoni,Harlander,LesHouches04}. 
In Table I we present the cross section of 
$pp \to ZA^0 \to \ell\bar{\ell} b\bar{b} +X$ via $b\bar{b} \to ZA^0$ in 
a two Higgs doublet model with $\tan\beta = 10$ and several values of $m_A$ 
for $\mu_F = M_Z +m_A, (M_Z+m_A)/2, (M_Z+m_A)/4$.
In our analysis, we have chosen $\mu_F = M_Z+m_A$ with the running bottom 
quark mass $m_b(m_A)$ in the Yukawa couplings. 
Our numerical values of cross section are comparable to the recent 
next-to-leading order QCD predictions for $Z^0 A^0$ associated production 
\cite{bbza-nlo}.

\begin{table}[h]
\caption{
The cross section in fb without cuts for 
$pp \to Z A^0 +X \to \ell\bar{\ell} b\bar{b} +X$ via bottom quark fusion 
($b\bar{b} \to Z A^0$) at $\sqrt{s} = 14$ TeV. 
We choose three values of the factorization scale ($\mu_F$) 
for $\tan\beta = 10$ and several values of $m_A$ 
in a two Higgs doublet model with $m_h = 120$ GeV, $m_H = m_A +100$ GeV 
and $\alpha_H = -\pi/4$.}
\centering\unskip\smallskip
\tabcolsep=1.5em
\begin{tabular}{c|ccccc}
\hline
$\mu_F \backslash  m_A$ (GeV) & 100 & 200 & 300 & 400 & 500 \\
\hline
$M_Z+m_A$     
& 28.3  & 1.81 & $8.72\times 10^{-2}$ & $9.19\times 10^{-3}$ 
& $1.92\times 10^{-3}$ \\
$(M_Z+m_A)/2$ 
& 22.3  & 1.52 & $7.58\times 10^{-2}$ & $8.19\times 10^{-3}$ 
& $1.74\times 10^{-3}$ \\
$(M_Z+m_A)/4$
& 16.1  & 1.19 & $6.25\times 10^{-2}$ & $6.99\times 10^{-3}$ 
& $1.53\times 10^{-3}$ \\
\hline
\end{tabular}
\end{table}

To study the effects of the Higgs scalar mixing angle ($\alpha_H$) in a two 
Higgs doublet model, we show the cross section of 
$pp \to Z A^0 +X \to \ell\bar{\ell}b\bar{b} +X$ versus $\alpha_H$
in Figure 3 for $\tan\beta =$ 2, 10, and 50 as well as 
(a) $m_A = 150$ GeV and (b) $m_A = 400$ GeV.
Also shown are the cross sections in the MSSM 
for $\tan\beta = 2$, 10, and 50.
In the MSSM, $\alpha_H$ becomes almost zero for $\tan\beta \sim 50$. 
For $\alpha_H < 0$, the cross section in a 2HDM is significantly larger 
than that in the MSSM.
We include contributions from both bottom quark fusion ($b\bar{b} \to ZA^0$) 
and gluon fusion ($gg \to ZA^0$).

For $m_A > m_Z +m_h$ and $m_A > m_t +m_W$, the branching fraction of 
$A^0 \to b\bar{b}$ is suppressed by $A^0 \to Zh^0$ and $A^0 \to t\bar{t}^*$ 
with real and virtual top quarks. 
Therefore, the cross section of $pp \to ZA^0 \to \ell\bar{\ell} b\bar{b} +X$ 
for $m_A = 400$ GeV is much smaller than that for $m_A = 150$ GeV. 

\section{THE PHYSICS BACKGROUND}
 
The dominant physics backgrounds to the final state of 
$ZA^0 \to \ell\bar{\ell}b\bar{b}$ come from $gg \to \ell\bar{\ell}b\bar{b}$, 
$q\bar{q} \to \ell\bar{\ell}b\bar{b}, \ell = e$ or $\mu$ 
and $pp \to W^+W^- b\bar{b} +X$. 
In our analysis, we actually evaluated 
$gg \to b\ell^+\nu \bar{b}\ell^-\bar{\nu}$ and 
$q\bar{q} \to b\ell^+\nu \bar{b}\ell^-\bar{\nu}$ with 
dominant contribution from $pp \to t\bar{t} \to bW^+\bar{b}W^- +X$.
In addition, we also consider backgrounds from 
$pp \to \ell\bar{\ell} gb +X$, $pp \to \ell\bar{\ell} g\bar{b} +X$, 
$pp \to \ell\bar{\ell} gq +X$, $pp \to \ell\bar{\ell} g\bar{q} +X$, and 
$pp \to \ell\bar{\ell} jj +X$, 
where $q = u, d, s$, or $c$ and $j = g, q$ or $\bar{q}$.
The programs MADGRAPH \cite{madgraph} and HELAS \cite{helas} are 
employed to evaluate the cross sections for all physics backgrounds. 

For an integrated luminosity ($L$) of 30 fb$^{-1}$, we require two isolated 
leptons with $p_T(\ell) > 15$ GeV and $|\eta(\ell)| < 2.5$ in each event. 
All jets are required to have $p_T(b,j) > 15$ GeV and $|\eta(b,j)| < 2.5$. 
The $b$-tagging efficiency ($\epsilon_b$) is taken to be $60\%$; 
the probability that a $c$-jet mistagged as a $b$-jet ($\epsilon_c$) 
is $10\%$, and the probability that any other jet mistagged as a $b$-jet 
($\epsilon_j$) is taken to be $1\%$. 
Furthermore, we require the invariant mass of the lepton pairs with opposite 
signs to be within 10 GeV of $M_Z$, 
that is $|M_{\ell\bar{\ell}}-M_Z| \le 10$~GeV 
to be the signature of a $Z$ boson.

For a higher integrated luminosity of 300 fb$^{-1}$, we require the same 
acceptance cuts as those for $L =$ 30 fb$^{-1}$, except $p_T(\ell) > $ 25 GeV 
and $p_T(b,j) > 30$ GeV. The $b$-tagging efficiency ($\epsilon_b$) is taken 
to be $50\%$, and the probability that a $c$-jet mistagged as a $b$-jet 
($\epsilon_c$) is $14\%$. 
We found that the $p_T$ cuts on leptons and bottom quarks are effective 
in removing most of the SM background, while most leptons from the $Z$ decays 
and most bottom quarks from the Higgs decays survive the $p_T$ cuts 
\cite{ggza2}. 

In addition, we require that the missing transverse energy ($\notE_T$) 
in each event should be less than 20 GeV for $L = 30$ fb$^{-1}$ 
and less than 40 GeV for $L = 300$ fb$^{-1}$. This cut on missing $E_T$ 
along with the requirement on the invariant mass of lepton pairs 
($|M_{\ell\bar{\ell}}-M_Z| \le 10$~GeV)
effectively reduce the background from $pp \to W^+W^- b\bar{b} +X$ which 
receives the major contribution from $pp \to t\bar{t} +X$.
Our acceptance cuts and efficiencies of $b$-tagging and mistagging 
are similar to those of the ATLAS collaboration \cite{ATLAS2}. 
 
\section{THE DISCOVERY POTENTIAL AT THE LHC}

To study the discovery potential of 
$pp \to Z A^0 \to \ell\bar{\ell} b\bar{b} +X$ at the LHC, 
we calculate the background from the SM processes of 
$pp \to \ell\bar{\ell} b\bar{b} +X$ in the mass window of
$m_A \pm \Delta M_{b\bar{b}}$ with $\Delta M_{b\bar{b}} = 22$ GeV.

We consider the Higgs signal to be observable 
if the $N \sigma$ lower limit on the signal plus background is larger than 
the corresponding upper limit on the background \cite{HGG,Brown}, namely,
\begin{eqnarray}
L (\sigma_S+\sigma_B) - N\sqrt{ L(\sigma_S+\sigma_B) } > 
L \sigma_B +N \sqrt{ L\sigma_B }
\end{eqnarray}
which corresponds to
\begin{eqnarray}
\sigma_S > \frac{N^2}{L} \left[ 1+2\sqrt{L\sigma_B}/N \right] \, .
\end{eqnarray}
Here $L$ is the integrated luminosity, 
$\sigma_S$ is the cross section of the Higgs signal, 
and $\sigma_B$ is the background cross section 
within a bin of width $\pm \Delta M_{b\bar{b}}$ centered at $m_A$. 
In this convention, $N = 2.5$ corresponds to a 5$\sigma$ signal.

In CP-conserving two Higgs doublet models with $m_A \gg M_Z$, 
a CP-even neutral Higgs boson with Standard-Model-like couplings may be the 
lightest scalar. In this decoupling limit \cite{decoupling}, 
$\sin^2(\beta-\alpha_H) \to 1$ and $\cos^2(\beta-\alpha_H) \to 0$.
We show the cross section with acceptance cuts in Figure 4 for 
$pp \to ZA^0 \to \ell\bar{\ell} b\bar{b} +X$ 
in a 2HDM with $m_H = m_A + 100$ GeV, $m_h = 120$ GeV, 
and $\alpha_H = \beta-\pi/2$ as well as the cross section in the MSSM, 
for $L = 30$ fb$^{-1}$ and $L = 300$ fb$^{-1}$.
The curves for the 5$\sigma$ and 3$\sigma$ cross sections 
for the $ZA^0$ signal are also presented.
We include contributions from both bottom quark fusion 
and gluon fusion.
With a luminosity of 30 fb$^{-1}$, it is possible to establish a 
5$\sigma$ signal of $ZA^0 \to \ell\bar{\ell} b\bar{b}$ 
for $m_A \alt 200$ GeV and $\tan\beta \sim 2$ or $\tan\beta \sim 50$. 
At a higher luminosity of 300 fb$^{-1}$ the discovery potential 
of this channel is greatly improved for $m_A \alt 280$ GeV 
and $\tan\beta \sim 2$ or $\tan\beta \sim 50$.
In the MSSM, it is difficult to observe the Higgs signal of $ZA^0$ 
since $m_H \sim m_A$.
In both models, if $m_A >$ 250 GeV and $\tan\beta \alt 7$, 
the branching fraction of $A^0 \to b\bar{b}$ 
is greatly suppressed when the Higgs pseudoscalar decays dominantly  
into $Zh^0$ and $t\bar{t}^*$ with one of the top quarks being virtual.

In Tables II and III, we present event rates after acceptance cuts for the 
Higgs signal ($N_S$) from $b\bar{b} \to ZA^0 \to \ell\bar{\ell} b\bar{b}$ 
and the background ($N_B$) as well as the ratio of signal to background 
$N_S/N_B$ and $N_S/\sqrt{N_B}$ 
in a two Higgs doublet model with $\tan\beta = 10$ and 50, 
$\alpha_H = -\pi/4$. 

\begin{table}[h]
\caption{
Event rates after acceptance cuts for the Higgs signal 
($N_S = \sigma_S \times L$) from $b\bar{b} \to ZA^0$ and the background 
($N_B = \sigma_B \times L$) as well as 
the ratio of signal to background $N_S/N_B$ and $N_S/\sqrt{N_B}$ 
in a two Higgs doublet model with $\tan\beta = 10$ and 50, 
$\alpha_H = -\pi/4$, and $m_H = m_A + 100$ GeV for an integrated luminosity 
of 30 fb$^{-1}$.}
\centering\unskip\smallskip
\tabcolsep=1.5em
\begin{tabular}{c|cccc}
\hline
$\tan\beta = 10$ \\
\hline
$m_A$ (GeV) & $N_S$ & $N_B$ & $N_S/N_B$ & $N_S/\sqrt{N_B}$ \\
\hline
100  & 196   & $1.01\times 10^4$ & 0.019     &  1.95    \\
200  &  14   &  2100             & 0.007     &  0.30    \\
300  &   1   &   577             & 0.001     &  0.03    \\
400  & $< 1$ &   193             & $< 0.001$ & $< 0.01$ \\
\hline
$\tan\beta = 50$ \\
\hline
$m_A$ (GeV) & $N_S$ & $N_B$ & $N_S/N_B$ & $N_S/\sqrt{N_B}$ \\
\hline
100  &  773  & $1.01\times 10^4$ & 0.076     &  7.7  \\
200  &  138  &  2100             & 0.066     &  3.0  \\
300  &   31  &   577             & 0.054     &  1.3  \\
400  &    9  &   193             & 0.045     &  0.62 \\
\hline
\end{tabular}
\end{table}

\begin{table}[h]
\caption{
The same as in Table I, 
except that the integrated luminosity is 300 fb$^{-1}$.}
\centering\unskip\smallskip
\tabcolsep=1.5em
\begin{tabular}{c|cccc}
\hline
$\tan\beta = 10$ \\
\hline
$m_A$ (GeV)  & $N_S$ & $N_B$ & $N_S/N_B$ & $N_S/\sqrt{N_B}$ \\
\hline
100  &  897  & $2.62\times 10^4$ & 0.034     &  5.5      \\
200  &   75  & $1.02\times 10^4$ & 0.007     &  0.74     \\
300  &    4  &   3380            & 0.001     &  0.06     \\
400  & $< 1$ &   1170            & $< 0.001$ &  0.01     \\
\hline
$\tan\beta = 50$ \\
\hline
$m_A$ (GeV) & $N_S$ & $N_B$ & $N_S/N_B$ & $N_S/\sqrt{N_B}$ \\
\hline
100  & 3310  & $2.62\times 10^4$ & 0.13     &  20.5 \\
200  &  731  & $1.02\times 10^4$ & 0.07     &   7.2 \\
300  &  168  &   3380            & 0.05     &   2.9 \\
400  &   46  &   1170            & 0.04     &   1.4 \\
\hline
\end{tabular}
\end{table}

The discovery contours for $pp \to ZA^0 \to \ell\bar{\ell} b\bar{b} +X$ via 
$b\bar{b} \to ZA^0$ at the LHC are presented in Figure 5 for an integrated 
luminosity of (a) $L = 30$ fb$^{-1}$ and (b) $L = 300$ fb$^{-1}$. 
We show 5$\sigma$ contours in the $(\alpha_H,\tan\beta)$ plane for 
$m_A = 150, 250$ and 400 GeV in a general two Higgs doublet model 
with $m_h = 120$ GeV and $m_H = m_A +100$ GeV. In addition, we present 
the curve for the decoupling limit with $\beta-\alpha_H = \pi/2$. 
For $L = 30$ fb$^{-1}$, it will be possible to discover the $ZA^0$ signal 
for $|\alpha_H| \alt 0.5$ and $m_A \alt 250$ GeV.
The higher luminosity ($L = 300$ fb$^{-1}$) greatly improves the reach for
$m_A$ up to 400 GeV in a large region of the parameter space 
with $|\alpha_H| \alt 1$.

\section{CONCLUSIONS}
 
Bottom quark fusion and gluon fusion are complementary processes 
to produce a Higgs pseudoscalar ($A^0$) and a $Z$ boson at the LHC.
While gluon fusion is the major source of $ZA^0$ for $\tan\beta \alt 10$, 
bottom quark fusion can make dominant contributions to the production 
of $ZA^0$ at the LHC for $\tan\beta \agt 10$.

We have found promising results for 
$pp \to ZA^0 \to \ell\bar{\ell}b\bar{b} +X$ via $b\bar{b} \to ZA^0$ 
in two Higgs doublet models at the LHC with $L = 300$ fb$^{-1}$ for 
$m_A \alt 400$ GeV, $\tan\beta \agt 5$, $|\alpha_H| \alt 1$, 
and $m_H = m_A +100$ GeV. 
In the MSSM with $m_A \agt 125$ GeV, $m_A \sim m_H$, and 
the production cross section of $gg \to ZA^0$ is usually small.  

Gluon fusion ($gg \to ZA^0$) offers great promise for $m_A \alt 260$ GeV and 
$\tan\beta \sim 2$ \cite{ggza2}.
The production rate of $ZA^0$ from gluon fusion at the LHC is
suppressed by the destructive interference between the triangle
and the box diagrams as well as the negative interference between
the top quark and the bottom quark loops, especially when they are comparable 
with $\tan\beta \sim 7$ \cite{ggza1}.

In a general two Higgs doublet model, the cross section of $b\bar{b} \to ZA^0$ 
and $gg \to ZA^0$ can be greatly enhanced when the heavier Higgs scalar 
($H^0$) can decay into the Higgs pseudoscalar and a $Z$ boson. 
If we take $m_H \sim m_A$ in a 2HDM, the Higgs signal will be reduced 
to the level of the MSSM. 
This discovery channel might provide a good opportunity to discover two Higgs 
bosons simultaneously if the heavier Higgs scalar ($H^0$) can decay into 
a $Z$ boson and a Higgs pseudoscalar ($A^0$).

\section*{ACKNOWLEDGMENTS}

We are grateful to Sally Dawson and Michael Spira for beneficial discussions.
This research was supported in part by the U.S. Department of Energy 
under grants 
No.~DE-FG02-04ER41305 and No.~DE-FG02-03ER46040.
 
\newpage

\begin{center}
{\large REFERENCES}
\end{center}
%


\newpage

\begin{figure}[htb]
\centering\leavevmode
\epsfxsize=6in
\epsfbox{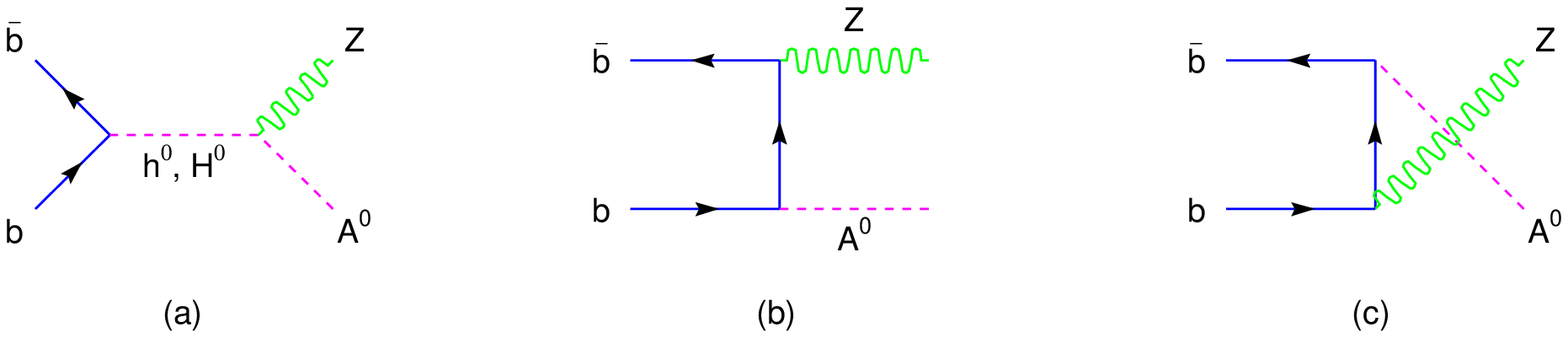}
\caption[]{
Feynman diagrams for the signal from $b\bar{b} \to Z A^0$.
\label{fig:Diagrams}
}\end{figure}

\newpage

\begin{figure}[htb]
\centering\leavevmode
\epsfxsize=6in
\epsfbox{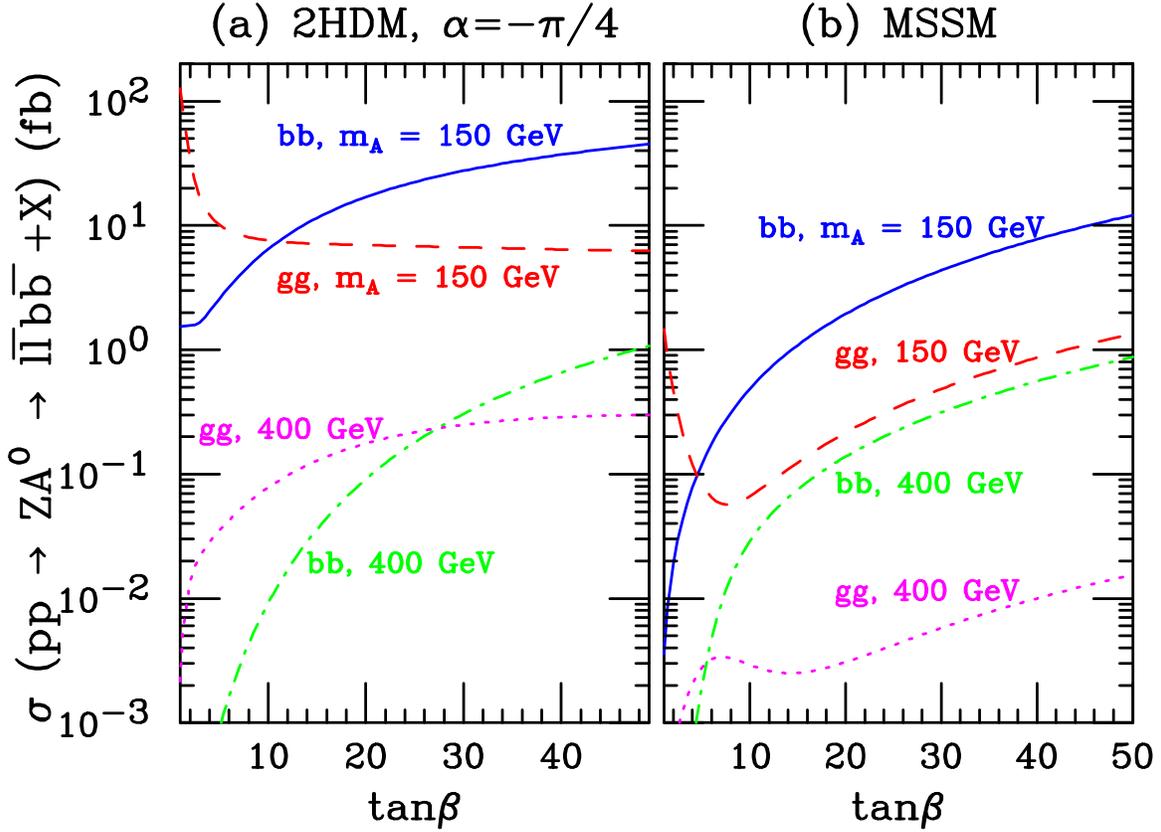}
\caption[]{
The cross section in fb without cuts for 
$pp \to Z A^0 +X \to \ell\bar{\ell} b\bar{b} +X$ 
at $\sqrt{s} = 14$ TeV, as a function of $\tan\beta$, 
for $m_A = 150$ and $400$ GeV, 
in (a) a two Higgs doublet model with $m_h = 120$ GeV, $m_H = m_A +100$ GeV 
and $\alpha_H = -\pi/4$
as well as 
in (b) the MSSM with $m_{\tilde{q}} = m_{\tilde{g}} = \mu = 1$ TeV. 
We show contributions from bottom quark fusion  
($b\bar{b} \to Z A^0$) and gluon fusion ($gg \to Z A^0$) separately.
\label{fig:tanb}
}\end{figure}


\begin{figure}[htb]
\centering\leavevmode
\epsfxsize=6in
\epsfbox{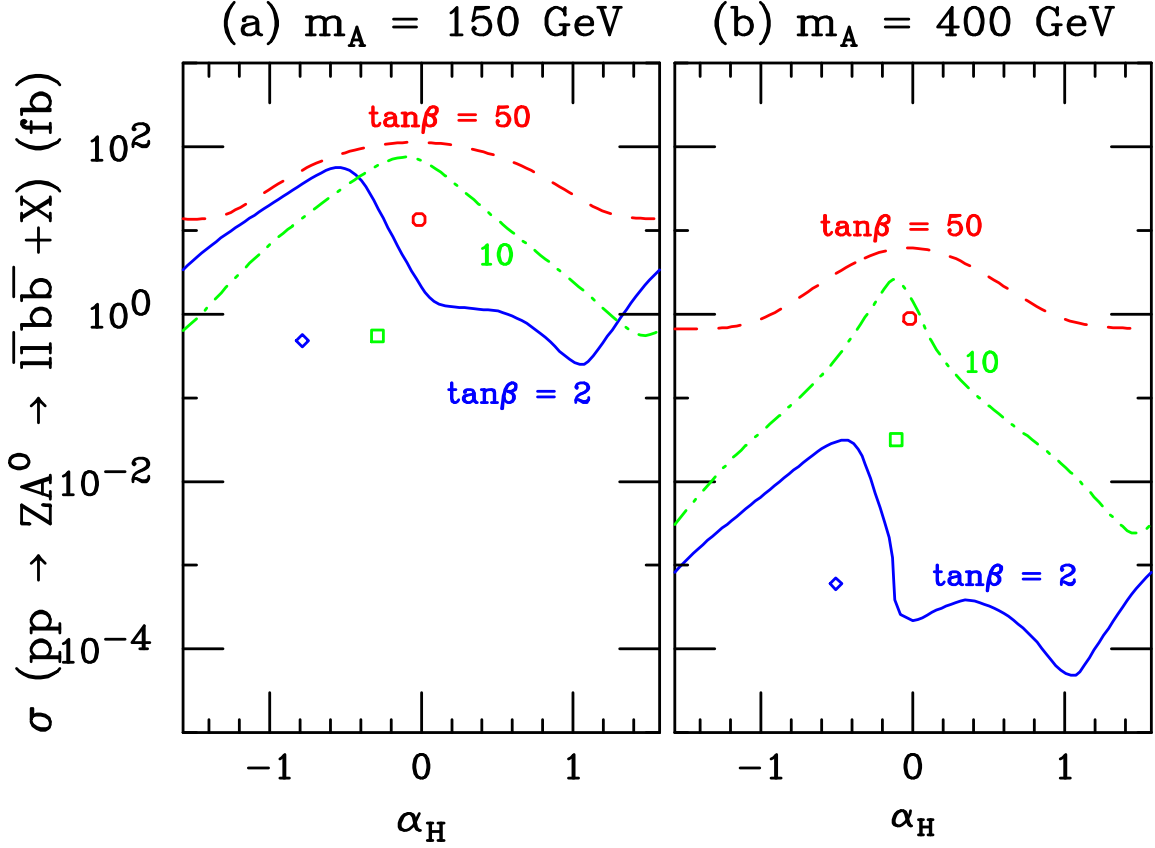}
\caption[]{
The cross section in fb without cuts 
for $pp \to Z A^0 +X \to \ell\bar{\ell}b\bar{b} +X$ at $\sqrt{s} = 14$ TeV, 
as a function of the Higgs scalar mixing angle $\alpha_H$,  
in a two Higgs doublet model with $m_h = 120$ GeV, $m_H = m_A +100$ GeV 
with $\tan\beta =$ 2, 10, and 50, for
(a) $m_A = 150$ GeV and (b) $m_A = 400$ GeV.
Also shown are the cross sections in the MSSM 
for $\tan\beta = 2$ (diamond), 10 (square), and 50 (circle).
We include contributions from bottom quark fusion 
($b\bar{b} \to Z A^0$) and gluon fusion ($gg \to Z A^0$).
\label{fig:alpha_H} 
}\end{figure}


\begin{figure}[htb]
\centering\leavevmode
\epsfxsize=6in
\epsfbox{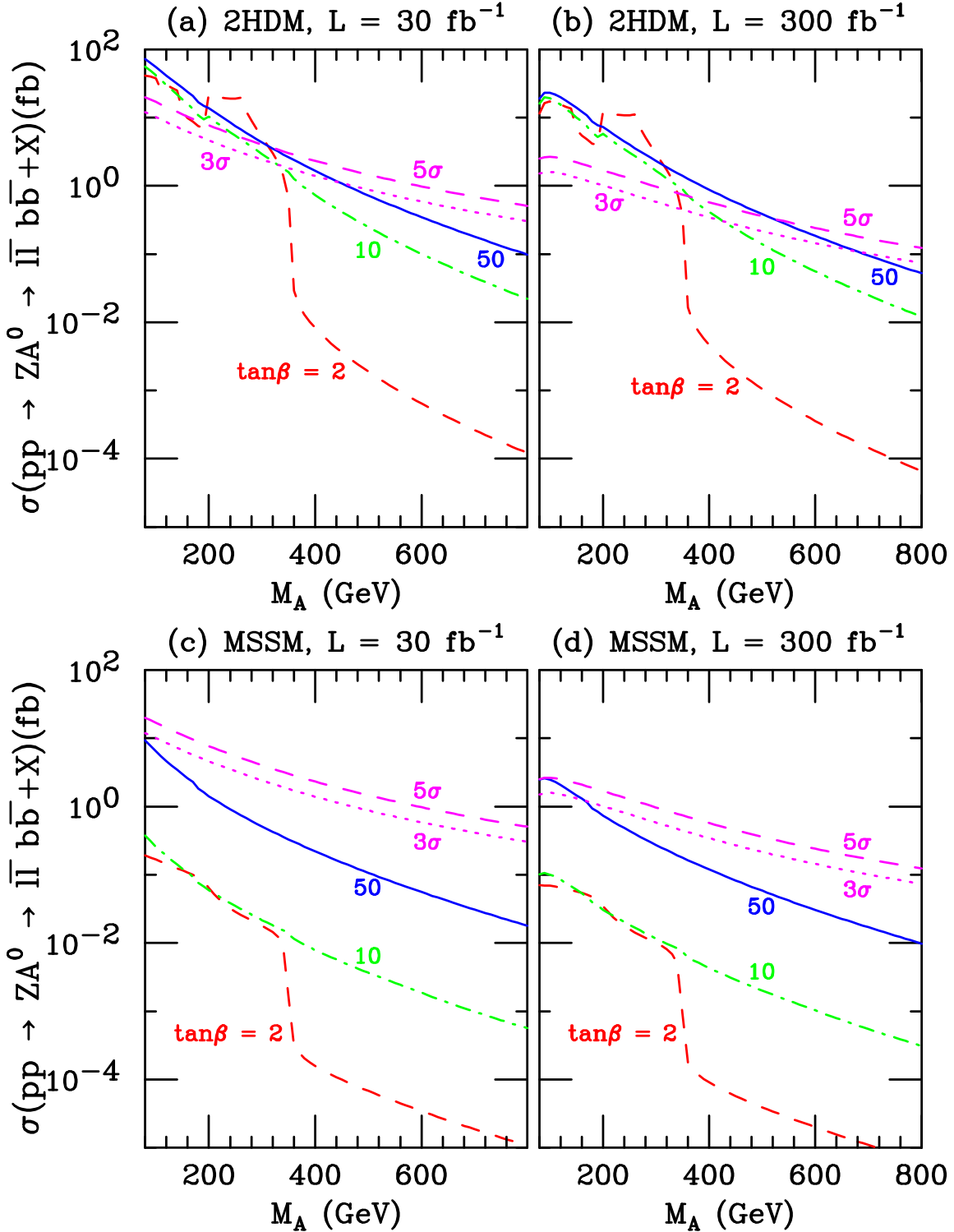}
\caption[]{
The cross section in fb for $pp \to Z A^0 +X \to \ell\bar{\ell} b\bar{b} +X$ 
versus $m_A$ at $\sqrt{s} = 14$ TeV, 
in a two Higgs doublet model with $m_h = 120$ GeV, $m_H = m_A +100$ GeV 
and $\alpha_H = \beta -\pi/2$ (the decoupling limit), 
for $\tan\beta = 2$ (dashed), 10 (dot-dashed), and 50 (solid). 
Also shown are the 5$\sigma$ (dashed) and 3$\sigma$ (dotted)  
cross sections for the $ZA^0$ signal required 
for an integrated luminosity~($L$) of (a) 30 fb$^{-1}$ and (b) 300 fb$^{-1}$. 
We have applied the acceptance cuts as well as 
the tagging and mistagging efficiencies described in the text.
\label{fig:discovery}
}\end{figure}

\newpage

\begin{figure}[htb]
\begin{minipage}[c]{\textwidth}
\begin{minipage}[b]{.49\textwidth}
\centering\leavevmode
\epsfxsize=3.4in
\epsfbox{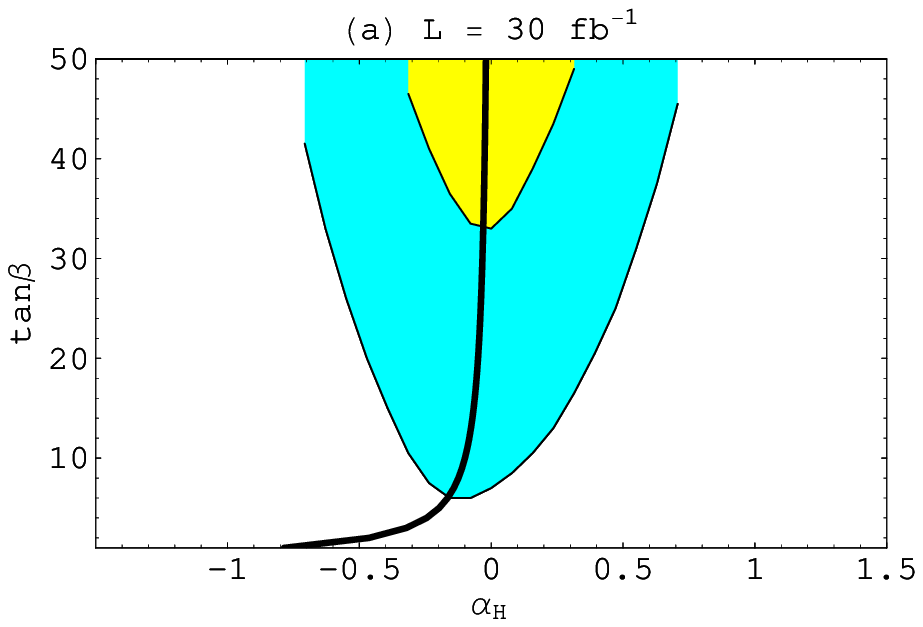}
\end{minipage}
\hfill
%
\begin{minipage}[b]{.49\textwidth}
\centering\leavevmode
\epsfxsize=3.4in
\epsfbox{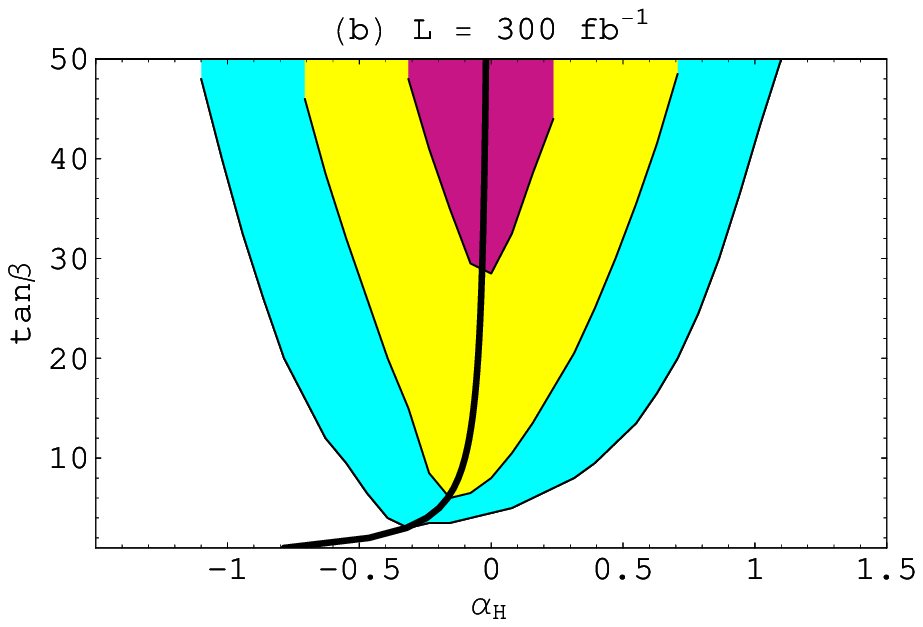}
\end{minipage}
\end{minipage}
\caption[]{
The $5\sigma$ discovery contours at the LHC 
with an integrated luminosity ($L$) of (a) 30 fb$^{-1}$ and (b) 300 fb$^{-1}$ 
in the ($\alpha_H$, $\tan\beta$) plane for $m_A = 150$ GeV (medium shading), 
$m_A = 250$ GeV (light shading), and $400$ GeV (dark shading) 
in a two Higgs doublet model with $m_h = 120$ GeV and $m_H = m_A +100$ GeV. 
The discovery region is the part of the parameter space above the contours.
In addition, we present the curve for the decoupling limit 
with $\beta-\alpha_H = \pi/2$. 
The Higgs signal includes contributions from $b\bar{b} \to Z A^0$ alone. 
We have applied the acceptance cuts as well as 
the tagging and mistagging efficiencies described in the text.
\label{fig:contour}
}\end{figure}

\end{document}